\documentclass[12pt]{iopart}

\usepackage{iopams}
\usepackage{graphicx}
\usepackage{color}
\usepackage[normalem]{ulem}

\newtheorem{proposition}{Proposition}
\newtheorem{theorem}[proposition]{Theorem}
\newtheorem{lemma}[proposition]{Lemma}

\newcommand{\CMP}{{\it {Commun.~Math.~Phys.~}}}

\newcommand{\bmath}[1]{\mbox{{\boldmath{{$#1$}}}}}

\begin{document}

\title[Existence of Einstein-Yang-Mills dyons]{On the existence of dyons and dyonic black holes in Einstein-Yang-Mills theory}

\author{Brien C. Nolan$^1$ and Elizabeth Winstanley$^2$}

\address{$^1$ School of Mathematical Sciences, Dublin City University,
Glasnevin, Dublin 9, Ireland}

\address{$^2$ Consortium for Fundamental Physics, School of Mathematics and Statistics,
\\
The University of Sheffield, Hicks Building, Hounsfield Road, Sheffield. S3 7RH
United Kingdom}

\eads{\mailto{Brien.Nolan@dcu.ie}, \mailto{E.Winstanley@sheffield.ac.uk}}

\begin{abstract}
We study dyonic soliton and black hole solutions  of the ${\mathfrak {su}}(2)$ Einstein-Yang-Mills equations in asymptotically anti-de Sitter space.
We prove the existence of non-trivial dyonic soliton and black hole solutions in a neighbourhood of the trivial solution.
For these solutions the magnetic gauge field function has no zeros and we conjecture that at least some of these non-trivial solutions will be stable. The global existence proof uses local existence results and a non-linear perturbation argument based on the (Banach space) implicit function theorem.
\end{abstract}

\section{Introduction}
\label{sec:intro}

The study of soliton and black hole solutions of Einstein-Yang-Mills (EYM) theory has a long history, sparked by the discovery of four dimensional, asymptotically flat, spherically symmetric, solitons \cite{Bartnik} and black holes \cite{AFBHs} when the gauge group is ${\mathfrak {su}}(2)$ and the gauge field is purely magnetic.  Although these solutions are unstable \cite{Straumann}, solutions of EYM in asymptotically flat space have been widely studied (see, for example, \cite{Volkov1} for a review).

Changing the space-time asymptotics dramatically  changes the properties of EYM solutions.
In this paper we focus on four-dimensional, asymptotically anti-de Sitter (adS) space-times.
The first EYM solutions in adS were found for gauge group ${\mathfrak {su}}(2)$ and,
as in asymptotically flat space, spherically symmetric
black hole \cite{Winstanley1} and soliton
\cite{Bjoraker} solutions exist for a purely magnetic gauge field.
However, unlike the asymptotically flat solutions, at least some of these purely magnetic, asymptotically adS, solutions are stable under linear perturbations
\cite{Winstanley1,Bjoraker,Sarbach}.
Similar results hold if the gauge group is enlarged to ${\mathfrak {su}}(N)$ \cite{Baxter,Winstanley}, yielding black holes with abundant gauge field hair \cite{Shepherd}.

As well as the existence of stable solitons and black holes with a purely magnetic gauge field, ${\mathfrak {su}}(2)$ EYM solutions in adS have another surprising feature: the existence (in numerical simulations) of spherically symmetric dyonic solutions for which the gauge field has non-trivial electric and magnetic parts \cite{Bjoraker}.  This is in contrast to the situation in asymptotically flat space-time, where the only EYM solution where the gauge field has a non-zero electric part is the trivial embedded Reissner-Nordstr\"om solution \cite{Ershov}.

In this paper we prove analytically the existence of four dimensional, spherically symmetric, dyonic soliton and black hole solutions of ${\mathfrak {su}}(2)$ EYM in adS, as previously found numerically \cite{Bjoraker}.  In section~\ref{sec:EYM} we briefly summarize our model, the field equations and boundary conditions. We also review the key features of the numerical solutions \cite{Bjoraker}.
We begin our analytic work in section~\ref{sec:local}, by proving the local existence of solutions of the field equations near the origin, event horizon and infinity.  In section~\ref{sec:global}, we then prove the existence of global solutions of the field equations, using a non-linear perturbation argument based on the implicit function theorem rather than the usual ``shooting'' type methods which were employed in, for example, \cite{Winstanley1}.  Our conclusions are presented in section~\ref{sec:conc}.

\section{${\mathfrak {su}}(2)$ Einstein-Yang-Mills theory}
\label{sec:EYM}

In this section we present the field equations for ${\mathfrak {su}}(2)$ Einstein-Yang-Mills theory in adS, and briefly review the properties of the numerical dyonic solutions \cite{Bjoraker}.

\subsection{Ansatz and field equations}
\label{sec:ansatz}

We consider four-dimensional EYM theory with a negative cosmological constant
$\Lambda <0$, with the following action:
\begin{equation}
S_{\mathrm {EYM}} = \frac {1}{2} \int d ^{4}x {\sqrt {-g}} \left[
R - 2\Lambda
- \Tr \, F_{\mu \nu }F ^{\mu \nu }
\right] ,
\label{eq:action}
\end{equation}
where $R$ is the Ricci scalar, $\Lambda $ the cosmological
constant, $F_{\mu\nu}$ the Yang-Mills gauge field and $\Tr $ denotes a Lie algebra trace.
We have fixed the gauge coupling constant $g=1$.
Throughout this paper, the metric has signature $\left( -, +, +, + \right) $ and
we use units in which $4\pi G = c = 1$.

The field equations derived from varying the action (\ref{eq:action}) are:
\begin{eqnarray}
T_{\mu \nu }  =  R_{\mu \nu } - \frac {1}{2} R g_{\mu \nu } +
\Lambda g_{\mu \nu },
\nonumber \\
0  =  D_{\mu } F_{\nu }{}^{\mu } = \nabla _{\mu } F_{\nu }{}^{\mu }
+ \left[ A_{\mu }, F_{\nu }{}^{\mu } \right] ,
\label{eq:fieldeqns}
\end{eqnarray}
where the YM stress-energy tensor is
\begin{equation}
T_{\mu \nu } =
\Tr F_{\mu \lambda } F_{\nu }{}^{\lambda }
- \frac {1}{4} g_{\mu \nu }
\Tr F_{\lambda \sigma} F^{\lambda \sigma } ,
\label{eq:Tmunu}
\end{equation}
and $F_{\mu \nu }$ is given in terms of the gauge potential $A_{\mu }$ by
\begin{equation}
F_{\mu \nu } = \partial _{\mu }A_{\nu } - \partial _{\nu }A_{\mu } +
\left[ A_{\mu },A_{\nu } \right] .
\label{eq:Fmunu}
\end{equation}
In (\ref{eq:Fmunu}), square brackets denote the Lie algebra commutator.

Considering static, spherically symmetric space-times, the metric in standard
Schwarzschild-like co-ordinates takes the form
\begin{equation}
ds^{2} = - \mu (r) S(r)^{2} \, dt^{2} + \mu (r)^{-1} \, dr^{2} +
r^{2} \, d\theta ^{2} + r^{2} \sin ^{2} \theta \, d\varphi ^{2} ,
\label{eq:metric}
\end{equation}
where the metric functions $\mu (r)$ and $S(r)$ depend on the radial co-ordinate
 $r$ only.
 The metric function $\mu (r)$ can be written as
\begin{equation}
\mu (r) = 1 - \frac {2m(r)}{r} + \frac {r^{2}}{\ell ^{2}},
\label{eq:mu}
\end{equation}
where the adS radius of curvature $\ell $ is defined by
\begin{equation}
\ell ^{2} = -\frac {3}{\Lambda }.
\end{equation}

With a suitable choice of gauge, the ${\mathfrak {su}}(2)$ gauge field takes the form
\cite{Kunzle}
\begin{equation}
A = {\mathcal {A}} \, dt + \frac {1}{2} \left( C - C^{H} \right) \, d \theta
- \frac {i}{2} \left[ \left( C + C^{H} \right) \sin \theta + D\cos \theta \right] \, d \varphi ,
\label{eq:gauge}
\end{equation}
where ${\mathcal {A}}$,  $C$ and $D$ are $2\times 2$ matrices, given by
\begin{equation}
{\mathcal {A}}
= \frac {i}{2} \left(
\begin{array}{cc}
\alpha (r) & 0 \\
0 & -\alpha (r)
\end{array}
\right) ,
\quad
C = \left(
\begin{array}{cc}
0 & \omega (r) \\
0 & 0
\end{array}
\right) ,
\quad
D = \left(
\begin{array}{cc}
1 & 0 \\
0 & -1
\end{array}
\right) ,
\end{equation}
with $\alpha (r)$ and $\omega (r)$ being real functions of $r$ only.
We emphasize that our gauge field ansatz (\ref{eq:gauge}) has both an electric part (${\mathcal {A}}\, dt$) described by the single function $\alpha (r)$ and a magnetic part described by the single function $\omega (r)$.  We will refer to $\alpha (r)$ as the ``electric gauge field function'' and $\omega (r)$ as the ``magnetic gauge field function''.

Substituting the metric (\ref{eq:metric}) and gauge field (\ref{eq:gauge}) into the field equations (\ref{eq:fieldeqns}) we obtain two Einstein equations:
\begin{eqnarray}
m'  =
\frac {r^{2}\alpha '^{2}}{2S^{2}} + \frac {\alpha ^{2} \omega ^{2}}{\mu S^{2}}
+ \mu \omega '^{2} + \frac {\left( \omega ^{2} -1 \right) ^{2}}{2r^{2}},
\nonumber
\\
\frac {S'}{S}  =  \frac {2\alpha ^{2} \omega ^{2}}{r\mu ^{2} S^{2}}
+ \frac {2\omega '^{2}}{r},
\label{eq:Eeqns}
\end{eqnarray}
and two Yang-Mills equations:
\begin{eqnarray}
\alpha ''  =
-\frac {2\alpha '}{r} + \frac {\alpha ' S'}{S}
+\frac {2\alpha \omega ^{2}}{r^{2} \mu } ,
\nonumber
\\
\omega ''  =
-\frac {\omega ' S'}{S} - \frac {\omega ' \mu '}{\mu }
- \frac {\alpha ^{2}\omega }{\mu ^{2}S^{2}}
+ \frac {\omega \left( \omega ^{2}-1 \right) }{r^{2} \mu }.
\label{eq:YMeqns}
\end{eqnarray}
In the above equations, a prime $'$ denotes $d/dr $.
The field equations (\ref{eq:Eeqns}--\ref{eq:YMeqns}) are invariant under the transformations $\alpha \rightarrow -\alpha $ and $\omega \rightarrow -\omega $ separately.
They are also invariant under the following scaling symmetry:
\begin{equation}
S(r) \rightarrow \lambda S(r), \qquad \alpha (r) \rightarrow \lambda \alpha (r),
\label{eq:scaling}
\end{equation}
for any constant $\lambda $. This corresponds to the invariance of the form (\ref{eq:metric}) of the metric and the form (\ref{eq:gauge}) of the gauge potential under scalings $t\rightarrow \lambda^{-1}t$.

Finally we note that, unlike the situation for purely magnetic ${\mathfrak {su}}(2)$ EYM \cite{Winstanley1,Bjoraker}, here the second Einstein equation for the metric function $S(r)$ does not decouple from the other equations.

\subsection{Boundary conditions}
\label{sec:boundary}

The field equations (\ref{eq:Eeqns}--\ref{eq:YMeqns}) have three singular points of interest\footnote{There may in addition be singular points corresponding to zeroes of $S$. However the second Einstein equation (\ref{eq:Eeqns}) is a first order, linear inhomogeneous equation in $S^2$ showing that, when solutions of the full system exist, $(S^2)^\prime\geq0$. We will be interested in solutions with a positive initial value for $S$, and so $S=0$ cannot arise. This is proven more carefully below.}, corresponding to the origin $r=0$, black hole event horizon $r=r_{h}$ (defined by zeroes of $\mu$, if there are any) and infinity $r\rightarrow \infty$.
In this section we state the boundary conditions on the functions $\alpha $,
$\omega  $, $\mu $ and $S$ at each of these singular points, before proving in
section~\ref{sec:local} that solutions of the field equations satisfying these boundary conditions exist.

\subsubsection{Origin}
\label{sec:origin}

We start by assuming that the field variables have regular Taylor series expansions about $r=0$:
\begin{eqnarray}
m(r)  =  m_{0}+m_{1}r+m_{2}r^{2}+m_{3}r^{3}+O( r^{4} ),
\nonumber \\
S(r)  =  S_{0} + S_{1}r + S_{2}r^{2} + O( r^{3}) ,
\nonumber \\
\alpha (r)  =  \alpha _{0} + \alpha _{1} r + \alpha _{2} r^{2} + \alpha _{3} r^{3}
+ O( r^{4} ) ,
\nonumber \\
\omega (r)  =  \omega _{0} + \omega _{1} r + \omega _{2} r^{2}+ O( r^{3} ) .
\label{eq:origin1}
\end{eqnarray}
To avoid a singularity either in the metric (\ref{eq:metric}) or the field equations
(\ref{eq:Eeqns}--\ref{eq:YMeqns}), it must be the case that
\begin{equation}
m_{0} = m_{1} = m_{2} = S_{1}=\alpha _{0} = \alpha _{2} =\omega _{1}=0
\end{equation}
and $\omega _{0}=\pm 1$.
We take $\omega _{0}=1$ without loss of generality since the field equations are invariant under the map $\omega \rightarrow -\omega $.
The constants $m_{3}$, $S_{2}$ and $\alpha _{3}$ are then determined in terms of
$S_{0}$, $\alpha _{1}$ and $\omega _{2}$ by the field equations and the series (\ref{eq:origin1}) become \cite{Bjoraker}:
\begin{eqnarray}
m(r)  =  \left( \frac {\alpha _{1}^{2}}{2S_{0}^{2}} + 2\omega _{2}^{2} \right)
r^{3}+O( r^{4} ),
\nonumber \\
S(r) =  S_{0} + \left( \frac {\alpha _{1}^{2}}{S_{0}} + 4S_{0} \omega _{2}^{2} \right) r^{2} + O( r^{3}) ,
\nonumber \\
\alpha (r)  =  \alpha _{1} r + \frac {\alpha _{1}}{5} \left(
\frac {2\alpha _{1}^{2}}{S_{0}^{2}} + 8 \omega _{2}^{2} + 2\omega _{2}
- \frac {1}{\ell ^{2}}
\right) r^{3}
+ O( r^{4} ) ,
\nonumber \\
\omega (r)  =  1 +\omega _{2} r^{2}+ O( r^{3} ) .
\label{eq:originfinal}
\end{eqnarray}
Subject to proving convergence, these series give us a three-parameter family of solutions near the origin, parameterized by $S_{0}\neq 0$, $\alpha _{1}$ and $\omega _{2}$ (plus the adS radius of curvature $\ell $).
In practise the value of $S_{0}$ is fixed by the requirement that $S(r)\rightarrow 1$ as $r\rightarrow \infty $.

\subsubsection{Event horizon}
\label{sec:horizon}

For a regular, non-extremal black hole event horizon at $r=r_{h}$, the value of
$m(r_{h})$ is fixed by the condition $\mu (r_{h})=0$ to be
\begin{equation}
m(r_{h}) = \frac {r_{h}}{2} + \frac {r_{h}^{3}}{2\ell ^{2}}.
\end{equation}
The first Yang-Mills equation (\ref{eq:YMeqns}) has a singularity at $r=r_{h}$ unless the electric gauge field function $\alpha $ vanishes there.
Series expansions of the field functions near the event horizon therefore take the form \cite{Bjoraker}:
\begin{eqnarray}
m (r)  =
\frac {r_{h}}{2} + \frac {r_{h}^{3}}{2\ell ^{2}} + m'_{h} \left( r - r_{h} \right)
+ O(r-r_{h})^{2},
\nonumber
\\
S (r)  =
S_{h} + S'_{h} \left( r- r_{h} \right) + O(r-r_{h})^{2},
\nonumber
\\
\alpha (r)  =
\alpha '_{h} \left( r- r_{h} \right) + O(r-r_{h})^{2},
\nonumber
\\
\omega (r)  =
\omega _{h}+ \omega '_{h} \left( r - r_{h} \right) + O(r-r_{h})^{2},
\label{eq:horizon}
\end{eqnarray}
where $m_{h}'$, $S_{h}'$ and $\omega _{h}'$ are given by the field equations
in terms of $\alpha _{h}'$ and $\omega _{h}$ as:
\begin{eqnarray}
m_{h}'  =  \frac {r_{h}^{2}\alpha _{h}'^{2}}{2S_{h}^{2}}+
\frac {\left( \omega _{h}^{2} -1 \right) ^{2}}{2r_{h}^{2}} ,
\nonumber
\\
S_{h}'  =
\frac {2\alpha _{h}'^{2}\omega _{h}^{2}}{S_{h}r_{h}\mu '(r_{h})^{2}}
+ \frac {2\omega _{h}'^{2}S_{h}}{r_{h}} ,
\nonumber
\\
\omega _{h}'  =
\frac {\omega _{h}}{r_{h}^{2} \mu '(r_{h})} \left( \omega _{h}^{2} -1 \right) ,
\end{eqnarray}
and
\begin{equation}
\mu '(r_{h}) = \frac {1}{r_{h}}-\frac {2m_{h}'}{r_{h}}+\frac { 3r_{h}}{\ell ^{2}} >0.
\end{equation}
In this case we also anticipate the existence of a four-parameter family of solutions, with parameters $r_{h}$, $S_{h}$, $\alpha _{h}'$ and $\omega _{h}$ (as well as the adS radius of curvature $\ell $).
As with the expansions near the origin,
the value of $S_{h}$ will be fixed by the requirement that $S(r)\rightarrow 1$ as
$r\rightarrow \infty $.

\subsubsection{Infinity}
\label{sec:infinity}

As $r\rightarrow \infty $, we require our metric (\ref{eq:metric}) to approach adS, so that $S(r) \rightarrow 1$ and $m(r)\rightarrow M$ where $M$ is a constant which corresponds to the mass of the soliton or black hole.
Assuming that the field variables have regular Taylor series expansions near infinity, the second of the Einstein equations (\ref{eq:Eeqns}) implies that $S'(r) = O(r^{-5})$ as $r\rightarrow \infty $, so the series take the form
\begin{eqnarray}
m(r)  =  M - \frac {1}{r} \left(
\frac {d_{1}^{2}}{2} + \alpha _{\infty }^{2} \omega _{\infty }^{2} \ell ^{2}
+ \frac {c_{1}^{2}}{\ell ^{2}}
+ \frac {\left( \omega _{\infty }^{2}-1 \right) ^{2}}{2}
\right) + O(r^{-2}) ,
\nonumber
\\
S(r)  =  1 - \frac {1}{2r^{4}} \left(
\alpha _{\infty }^{2} \omega _{\infty }^{2} \ell ^{4} + c_{1}^{2}
\right)  + O(r^{-5}) ,
\nonumber
\\
\alpha (r)  =  \alpha _{\infty } + \frac {d_{1}}{r} + O(r^{-2}),
\nonumber
\\
\omega (r)  =  \omega _{\infty } + \frac {c_{1}}{r} + O(r^{-2}).
\label{eq:infinity}
\end{eqnarray}
The field equations place no restrictions on the constants $\alpha _{\infty }$,
$d_{1}$, $\omega _{\infty }$, $c_{1}$ or $M$, giving an expected five-parameter family of solutions  (which depend on the adS radius of curvature $\ell $ as well as the parameters above).

\subsection{Trivial solutions}
\label{sec:trivial}


The field equations (\ref{eq:Eeqns}--\ref{eq:YMeqns}) have a few simple solutions:
\begin{enumerate}
\item
If we set
\begin{equation}
\alpha (r) \equiv 0, \qquad \omega (r) \equiv \pm 1 ,
\label{eq:SadS}
\end{equation}
then both $m(r)$ and $S(r)$ are constants, giving the Schwarzschild-adS black hole solution.  If, in addition, $m(r) \equiv 0$ then pure adS is also a solution of the system.
\item
Alternatively, if we set
\begin{equation}
\alpha (r) \equiv 0, \qquad \omega (r) \equiv 0 ,
\end{equation}
then $S(r)$ is still constant, but in this case
\begin{equation}
m(r) = M - \frac {1}{2r},
\end{equation}
giving the metric for a magnetically charged, Abelian, Reissner-Nordstr\"om-adS
black hole with charge $Q_{M}=1$.
\item
The Abelian Reissner-Nordstr\"om-adS black hole with electric charge $Q_{E}$ and magnetic charge equal to unity is also a solution, obtained by setting
\begin{equation}
\alpha (r) = \frac {Q_{E}}{r}, \qquad \omega \equiv 0,
\end{equation}
in which case $S(r) \equiv 1$ and
\begin{equation}
m(r)  = M - \frac {\left( Q_{E}^{2}+1 \right) }{2r}.
\end{equation}
Due to the coupling between the electric and magnetic gauge field functions in the Yang-Mills equations (\ref{eq:YMeqns}), Abelian
Reissner-Nordstr\"om-adS with non-zero electric charge but zero magnetic charge
is not a solution of the system.
\item
Finally, by setting $\alpha (r) \equiv 0$, the field equations
(\ref{eq:Eeqns}--\ref{eq:YMeqns}) reduce to those for purely magnetic
${\mathfrak {su}}(2)$ EYM theory with a negative cosmological constant, whose solutions have already been studied in depth \cite{Winstanley1,Bjoraker,Winstanley}.
In this case the second Einstein equation for the metric function $S(r)$
(\ref{eq:Eeqns}) decouples from the other equations.
\end{enumerate}
Our focus in this paper is non-trivial solutions (that is, solutions which do not appear in the list above), in a neighbourhood of either the pure adS solution or Schwarzschild-adS solution.

\subsection{Properties of the numerical solutions}
\label{sec:numerical}

The dyonic field equations (\ref{eq:Eeqns}--\ref{eq:YMeqns}) were solved numerically in \cite{Bjoraker}, so here we briefly review those results.
The field equations are solved by starting the integration close to either the origin (for soliton solutions) or event horizon (for black hole solutions), using the relevant series ((\ref{eq:originfinal}) or (\ref{eq:horizon}) respectively) and integrating outwards to large $r$, stopping either once the field variables have converged to their asymptotic values within a desired numerical tolerance or the solutions have become singular.
To satisfy the boundary condition on $S(r)$ at infinity, it is simplest to initially set either $S_{0}=1$ (for soliton solutions) or $S_{h}=1$ (for black hole solutions) and integrate the field equations.  The value of $S(r)$ as $r\rightarrow \infty $ will then be $S_{\infty }$ which is in general not equal to unity.  Performing a scaling transformation (\ref{eq:scaling}) with $\lambda = S_{\infty }^{-1}$ then gives a numerical solution for which the boundary conditions (\ref{eq:infinity}) hold.
With this approach, the parameters which can be varied are $\alpha _{1}$ and
$\omega _{2}$ for soliton solutions and $r_{h}$, $\alpha _{h}'$ and $\omega _{h}$ for black hole solutions, as well as the  adS radius of curvature $\ell $.

\begin{figure}
\begin{center}
\includegraphics[width=10cm]{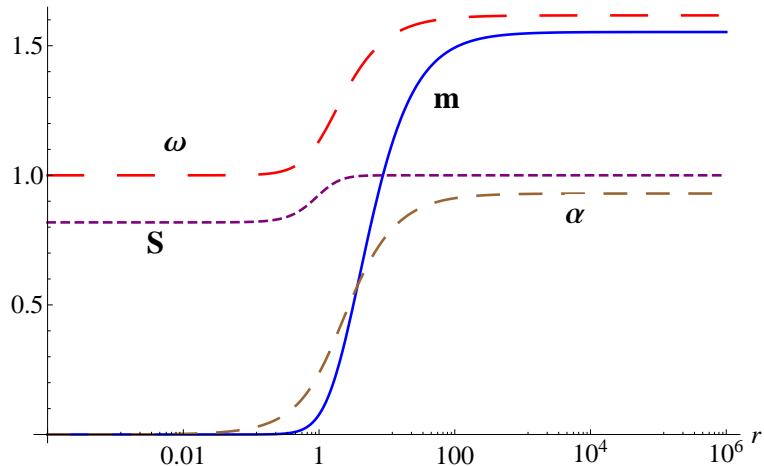}
\end{center}
\caption{Typical dyonic soliton solution with $\ell =1$, $\alpha _{1}=0.2455$
and $\omega _{2}=0.2$. Both the electric gauge field function
$\alpha (r)$ and the
magnetic gauge field function $\omega (r)$ are monotonically increasing and neither has any zeros for $r>0$.
}
\label{fig:one}
\end{figure}

\begin{figure}
\begin{center}
\includegraphics[width=10cm]{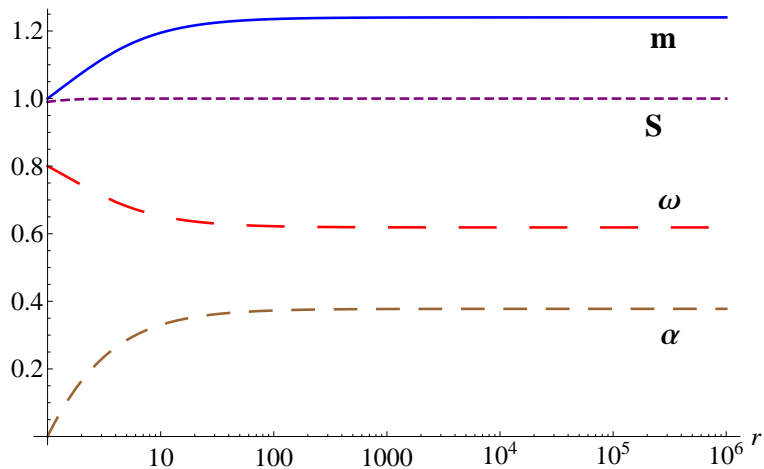}
\end{center}
\caption{Typical dyonic black hole solutions with $\ell =1$, $r_{h}=1$,
$\alpha _{h}'=0.2971$ and $\omega _{h}=0.8$. The electric gauge field function $\alpha (r)$ is monotonically increasing and the magnetic gauge field  function $\omega (r)$ is monotonically decreasing. Neither gauge function has any zeros for $r>r_{h}$.
}
\label{fig:two}
\end{figure}

For each value of $\ell $ and $r_{h}$ (with $r_{h}=0$ for solitons), regular soliton and black hole solutions are found in continous regions of the two-dimensional parameter space
$\left( \alpha _{1}, \omega _{2} \right)$ or
$\left( \alpha _{h}', \omega _{h} \right) $.
The size of these regions increase as the adS radius $\ell $ decreases (or, alternatively, as $\left| \Lambda \right|$ increases).
Some typical solutions are shown in figures~\ref{fig:one} and \ref{fig:two}.

It is straightforward to show from the field equations (\ref{eq:Eeqns}--\ref{eq:YMeqns}) that the electric gauge field function $\alpha (r)$ is monotonic, as are the metric functions $m(r)$ and $S(r)$.
This is not the case for the magnetic gauge field function $\omega (r)$, which typically has some zeros before approaching its asymptotic values.
However, we find solutions in which $\omega (r)$ has no zeros (that is, is nodeless)
in a neighbourhood of the trivial pure adS (for solitons) or Schwarzschild-adS
(for black holes) solution.
We observe that the size of the neighbourhood in which these nodeless solutions
exist shrinks as the adS radius of curvature $\ell $ increases (or
$\left| \Lambda \right| $ decreases).
Similar behaviour is observed in the purely magnetic case \cite{Winstanley1,Bjoraker}.
Further details of the space of solutions can be found in \cite{Bjoraker,Shepherd1}.

\section{Local existence}
\label{sec:local}

In this section we prove the existence of local families of solutions of the field equations (\ref{eq:Eeqns}--\ref{eq:YMeqns}) in neighbourhoods of the singular points $r=0$ (origin), $r=r_{h}$ (black hole event horizon) and $r\rightarrow \infty $ (infinity).
We follow the method of \cite{BFM}, making use of the following theorem:

\begin{theorem}
\label{thm:BFM}
\cite{BFM} Consider a system of differential equations for $n+m$ functions
${\bmath {u}}=\left( u_{1},u_{2},\ldots u_{n} \right) ^{T}$
and ${\bmath {v}} = \left( v_{1}, v_{2}, \ldots v_{m} \right) ^{T}$
of the form
\begin{eqnarray}
x \frac {du_{i}}{dx} = x^{\sigma _{i}} f_{i}\left(x,{\bmath {u}},{\bmath {v}}\right) ,
\nonumber \\
x \frac {dv_{i}}{dx} = - \tau _{i} v_{i} + x^{\varsigma _{i}} g_{i} \left( x, {\bmath {u}}, {\bmath {v}} \right) ,
\label{eq:BFMsys}
\end{eqnarray}
with constants $\tau _{i}>0$ and integers $\sigma _{i}, \varsigma _{i} \ge 1$, and let ${\cal {C}}$ be
an open subset of ${\mathbb {R}}^{n}$ such that the functions $f_{i}$, $i=1,\ldots n$ and $g_{i}$, $i=1,\ldots m$
are analytic in a neighbourhood of $x=0$, ${\bmath {u}}={\bmath {c}}$, ${\bmath {v}}={\bmath {0}}$
for all ${\bmath {c}} \in {\cal {C}}$.
Then there exists an $n$-parameter family of solutions of the system (\ref{eq:BFMsys})
such that
\begin{equation}
u_{i}(x) = c_{i} + O\left( x^{\sigma _{i}} \right) ,
\qquad
v_{i}(x) = O \left( x^{\varsigma _{i}} \right) ,
\end{equation}
where $u_{i}(x)$, $i=1,\ldots n$ and $v_{i}(x)$, $i=1,\ldots m$ are defined for ${\bmath {c}} \in {\cal {C}}$,
for $\left| x \right| < x_{0}({\bmath {c}})$ (for some $x_{0}({\bmath {c}})>0)$ and are analytic
in $x$ and ${\bmath {c}}$.
\end{theorem}

The key aspect of applying this theorem is setting the field equations in the required form (\ref{eq:BFMsys}) near each of the singular points.  Proving the local existence of families of solutions with the desired boundary conditions
(\ref{eq:originfinal}, \ref{eq:horizon}, \ref{eq:infinity}) then follows in a straightforward way from the above theorem.
One advantage of using Theorem~\ref{thm:BFM} is that the local solutions are analytic in the independent variable $x$ and also in the parameters ${\bmath {c}}$.
This is helpful when proving the global existence of solutions in
section~\ref{sec:global}.
In each case the proof of local existence is very similar to that in \cite{BFM}, although the presence of the electric gauge field function $\alpha (r)$ complicates the analysis both in terms of the boundary conditions at the origin and event horizon, and also because, unlike the situation for purely magnetic configurations, here we must include the field equation for the metric function $S(r)$.
For each of the three propositions in this section, we also add the equation
$d{\ell }/dx=0$ (for each particular choice of independent variable $x$), which clearly has the trivial solution $\ell = \ell_{0}$. Adding this equation to our system means that the local solutions whose existence we prove will also be analytic in $\ell _{0}$, or, equivalently, in $\ell $.

\begin{proposition}
\label{thm:origin}
 (Local existence of solutions in a neighbourhood of the origin)
 \\
 There exists a three-parameter family of local solutions of the field equations
(\ref{eq:Eeqns}--\ref{eq:YMeqns}) near $r=0$, satisfying the boundary conditions
(\ref{eq:originfinal}) and analytic in $S_{0}$, $\alpha _{1}$, $\omega _{2}$, $r$ and $\ell$.
\end{proposition}

\paragraph{{\bf {Proof}}}
Following \cite{BFM}, we take our independent variable $x=r$, and we define the following dependent variables:
\begin{eqnarray}
 \lambda _{1}(x) = \frac {1}{r^{2}}\left[ 1 - \mu (r) \right] , \qquad
\lambda _{2} = S(r),
\nonumber \\
\psi _{1}(x) = \frac {\alpha (r)}{r}, \qquad
\psi _{2}(x) = \frac {1}{r^{2}} \left[ \omega (r) - 1 \right] ,
\nonumber \\
\xi _{1}(x) = \frac {1}{r} \frac {d\alpha }{dr} - \frac {\alpha }{r^{2}},
\qquad
\xi _{2}(x) = \frac {\mu }{r} \frac {d\omega }{dr}.
\end{eqnarray}
These dependent variables are the same as in the corresponding proof in \cite{BFM}, apart from $\psi _{1}$ and $\xi _{1}$ which depend on the electric gauge field function
$\alpha (r)$.
In terms of these variables, the field equations (\ref{eq:Eeqns}--\ref{eq:YMeqns}) take the form
\begin{eqnarray}
x\frac {d\lambda _{1}}{dx}  =
-3\lambda _{1} - \frac {3}{\ell ^{2}}
+ \frac {3\psi _{1}^{2}}{\lambda _{2}^{2}} + 4 \psi _{2}^{2} + 2\xi _{2}^{2}
+ x{\mathcal {G}}_{0,1},
\nonumber
\\
x \frac {d\lambda _{2}}{dx}  =
x^{2}{\mathcal {G}}_{0,2},
\nonumber
\\
x\frac {d\psi _{1}}{dx}  =  x\xi _{1},
\nonumber \\
x\frac {d\psi _{2}}{dx}  =  -2\psi _{2} + \xi _{2}
 + x^{2}{\mathcal {J}}_{0,2},
 \nonumber \\
x\frac {d\xi _{1}}{dx}  =
-4\xi _{1} + x{\mathcal {H}}_{0,1},
\nonumber \\
x\frac {d\xi _{2}}{dx}  =
-\xi _{2} + 2\psi _{2} + x^{2} {\mathcal {H}}_{0,2} ,
\label{eq:originBFM1}
\end{eqnarray}
where ${\mathcal {G}}_{0,i}$ and ${\mathcal {H}}_{0,i}$ are polynomials in
the $\lambda _{i}$, $\left( 1- x^{2} \lambda _{1}\right) ^{-1}$, $\lambda _{2}^{-1}$,
the $\psi _{i}$ and the $\xi _{i}$ and are given explicitly by:
\begin{eqnarray}
{\mathcal {G}}_{0,1} &  = &
\frac {\xi _{1}}{\lambda _{2}^{2}} \left( 2\psi _{1} + x\xi _{1} \right)
+ x\psi _{2}^{3} \left( 4+ x^{2}\psi _{2} \right)
+ \frac {2x\lambda _{1}\xi _{2}^{2}}{1-x^{2} \lambda _{1}}
\nonumber \\
& & + \frac {2x\psi _{1}^{2}\left( \lambda _{1}+2\psi _{2}+x^{2} \psi _{2}^{2} \right)
}{\left( 1- x^{2}\lambda _{1} \right) \lambda _{2}^{2}} ,
\nonumber
\\
{\mathcal {G}}_{0,2} & = &
\frac {2\psi _{1}^{2}\left( 1 + x^{2} \psi _{2} \right) ^{2}}{\left(
1- x^{2} \lambda _{1} \right) ^{2} \lambda _{2} }
+ \frac {2\lambda _{2} \xi _{2}^{2}}{\left( 1 - x^{2} \lambda _{1}\right) ^{2}},
\nonumber \\
{\mathcal {H}}_{0,1} & = &
\frac {2\psi _{1} \left( \lambda _{1} + 2\psi _{2} + x^{2} \psi _{2}^{2} \right)
}{1-x^{2} \lambda _{1} }
+ \frac {\psi _{1} +x\xi _{1}}{\lambda _{2}}{\mathcal {G}}_{0,2},
\nonumber \\
{\mathcal {H}}_{0,2} & = &
-\frac {\psi _{1}^{2}\left( 1 + x^{2} \psi _{2} \right) }{\left(
1- x^{2} \lambda _{1} \right) \lambda _{2}^{2}}
+ \psi _{2}^{2} \left( 3 + x^{2} \psi _{2} \right)
- \frac {\xi _{2}}{\lambda _{2}} {\mathcal {G}}_{0,2},
\nonumber \\
{\mathcal {J}}_{0,2} & = &
\frac {\lambda _{1}\xi _{2}}{1 -x^{2} \lambda _{1}}.
\end{eqnarray}
A further transformation is needed to cast the equations (\ref{eq:originBFM1}) into the required form (\ref{eq:BFMsys}). In particular, we define new variables $u_{0}$, $v_{0}$ and ${\tilde {\lambda }}_{1}$ as follows \cite{BFM}:
\begin{eqnarray}
  u_{0} = \frac {1}{3} \left( \psi _{2} + \xi _{2} \right),
\qquad
v_{0} = \frac {1}{3} \left( 2\psi _{2} - \xi _{2} \right),
\nonumber \\
{\tilde {\lambda }}_{1} = \lambda _{1} + \frac {1}{\ell ^{2} }
- \frac {\psi _{1}^{2}}{\lambda _{2}^{2}}
- 4u_{0}^{2} + 2v_{0}^{2} ,
\end{eqnarray}
in terms of which the equations for $\lambda _{1}$, $\psi _{2}$ and $\xi _{2}$ in
(\ref{eq:originBFM1}) are replaced by
\begin{eqnarray}
x\frac {d{\tilde {\lambda }}_{1}}{dx} = -3{\tilde {\lambda }}_{1}
+ x {\tilde {{\mathcal {G}}}}_{0,1},
\qquad
x\frac {du_{0}}{dx} = x^{2}{\mathcal {Q}}_{0,1},
\nonumber \\
x\frac {dv_{0}}{dx} = -3v_{0} + x^{2} {\mathcal {Q}}_{0,2},
\end{eqnarray}
where
\begin{eqnarray}
{\tilde {{\mathcal {G}}}}_{0,1}  =
{\mathcal {G}}_{0,1} - \frac {2\psi _{1}\xi _{1}}{\lambda _{2}^{2}}
+ \frac {2x\psi _{1}^{2}}{\lambda _{2}^{3}}{\mathcal {G}}_{0,2}
-8xu_0 {\mathcal {Q}}_{0,1} + 4xv_0 {\mathcal {Q}}_{0,2},
\nonumber \\
{\mathcal {Q}}_{0,1}  =
\frac {1}{3} \left(
{\mathcal {J}}_{0,2} + {\mathcal {H}}_{0,2}  \right) ,
\qquad
{\mathcal {Q}}_{0,2}  =
\frac {1}{3} \left( 2{\mathcal {J}}_{0,2} - {\mathcal {H}}_{0,2} \right).
\end{eqnarray}
The field equations are now in the form to apply Theorem~\ref{thm:BFM}, giving, in a neighbourhood of $x=0$, a three-parameter family of solutions of the form
\begin{eqnarray}
{\tilde {\lambda }}_{1}=O(x), \qquad \lambda _{2} = S_{0} + O(x^{2}),
\qquad
\psi _{1} = \alpha _{1} + O(x),
\nonumber \\
\xi _{1} = O(x), \qquad
u_{0}=\omega _{2}+O(x^{2}), \qquad
v_{0}= O(x^{2}).
\end{eqnarray}
Restoring the original field variables, we have proven the existence of solutions of the field equations in a neighbourhood of the origin, satisfying the boundary conditions (\ref{eq:originfinal}) and analytic in $S_{0}$, $\alpha _{1}$, $\omega _{2}$, $r$ and $\ell $.
\hfill $\square $

\begin{proposition}
\label{thm:horizon}
(Local existence of solutions in a neighbourhood of the event horizon) \\
There exists a four-parameter family of local solutions of the field equations
(\ref{eq:Eeqns}--\ref{eq:YMeqns}) near an event horizon at $r=r_{h}$, satisfying the boundary conditions (\ref{eq:horizon}) and analytic in
$r_{h}$, $S_{h}$, $\alpha _{h}'$, $\omega _{h}$, $r$ and $\ell $.
\end{proposition}

\paragraph{{\bf {Proof}}}

Let $r_h>0$. Following \cite{BFM}, we take our independent variable $x=r-r_{h}$ and define the following dependent variables:
\begin{eqnarray}
\rho (x) = r,
\qquad
\lambda _{1}(x) = \frac {\mu (r)}{x},
\qquad
\lambda _{2}(x) = S(r),
\nonumber \\
\psi _{1}(x) = \frac {\alpha (r)}{x}, \qquad
\psi _{2}(x) = \omega (r),
\nonumber \\
\xi _{1}(x) = \frac {\rho (x)^{2}}{S(r)}\frac {d\alpha }{dr},
\qquad
\xi _{2}(x) = \frac {\mu (r)}{x} \frac {d\omega }{dr}.
\end{eqnarray}
Again, these variables are the same as those in \cite{BFM}, apart from $\lambda _{2}$, $\psi _{1}$ and $\xi _{1}$.  In terms of our new variables, the field equations
(\ref{eq:Eeqns}--\ref{eq:YMeqns}) take the form
\begin{eqnarray}
x\frac {d\rho }{dx} = x,
\nonumber \\
x \frac {d\lambda _{1}}{dx} =
-\lambda _{1} + {\mathcal {F}}_{h,1} + x{\mathcal {G}}_{h,1}
,
\nonumber \\
x \frac {d\lambda _{2}}{dx} = x{\mathcal {G}}_{h,2}
,
\nonumber \\
x\frac {d\psi _{1}}{dx} =
-\psi _{1} + \frac {\lambda _{2}\xi _{1}}{\rho ^{2}}
,
\nonumber \\
x\frac {d\psi _{2}}{dx} = \frac {x\xi _{2}}{\lambda _{1}}
,
\nonumber \\
x\frac {d\xi _{1}}{dx} =
\frac {2x\psi _{1}\psi _{2}^{2}}{\lambda _{1}\lambda _{2}}
,
\nonumber \\
x\frac {d\xi _{2}}{dx} = -\xi _{2} + {\mathcal {P}}_{h,2} + x{\mathcal {H}}_{h,2}
,
\end{eqnarray}
where
\begin{eqnarray}
{\mathcal {F}}_{h,1} =
\frac {1}{\rho } - \frac {1}{\rho ^{3}} + \frac {3\rho }{\ell ^{2}}
+ \frac {\psi _{2}^{2} \left( 2 - \psi _{2}^{2} \right) }{\rho ^{3}}
- \frac {\xi _{1}^{2}}{\rho ^{3}},
\nonumber \\
{\mathcal {G}}_{h,1} =
-\frac {\lambda _{1}}{\rho } - \frac {2\psi _{1}^{2} \psi _{2}^{2}}{\rho
\lambda _{1}\lambda _{2}^{2}}
- \frac {2\xi _{2}^{2}}{\rho \lambda _{1}},
\nonumber \\
{\mathcal {G}}_{h,2} =
\frac {2\psi _{1}^{2} \psi _{2}^{2}}{\rho \lambda _{1}^{2} \lambda _{2}}
+ \frac {2\lambda _{2}\xi _{2}^{2}}{\rho \lambda _{1}^{2}},
\nonumber \\
{\mathcal {P}}_{h,2} =
\frac {\psi _{2}\left( \psi _{2}^{2} - 1 \right) }{\rho ^{2}},
\nonumber \\
{\mathcal {H}}_{h,2} =
-\frac {\psi _{1}^{2}\psi _{2}}{\lambda _{1}\lambda _{2}^{2}}
- \frac {\xi _{2} {\mathcal {G}}_{h,2}}{\lambda _{2}}.
\end{eqnarray}
To achieve the form (\ref{eq:BFMsys}) required for application of
Theorem~\ref{thm:BFM}, we make the further variable transformation
\begin{equation}
{\tilde {\psi }}_{1} = \psi _{1} - \frac {\lambda _{2}\xi _{1}}{\rho ^{2}},
\qquad
{\tilde {\lambda }}_{1} = \lambda _{1} - {\mathcal {F}}_{h,1},
\qquad
{\tilde {\xi }}_{2} = \xi _{2} - {\mathcal {P}}_{h,2}.
\end{equation}
The modified field equations then take the form
\begin{eqnarray}
x\frac {d{\tilde {\lambda }}_{1}}{dx} =
-{\tilde {\lambda }}_{1} + x{\tilde {\mathcal {G}}}_{h,1},
\nonumber \\
x \frac {d{\tilde {\psi }}_{1}}{dx} =
- {\tilde {\psi }}_{1} + x{\mathcal {J}}_{h,1},
\nonumber \\
x\frac {d{\tilde {\xi }}_{2}}{dx} =
-{\tilde {\xi }}_{2} + x{\tilde {\mathcal {H}}}_{h,2},
\end{eqnarray}
where
\begin{eqnarray}
{\tilde {\mathcal {G}}}_{h,1}=
{\mathcal {G}}_{h,1} - \frac {\partial {\mathcal {F}}_{h,1}}{\partial \rho }
- \frac {\xi _{2}}{\lambda _{1}}
\frac {\partial {\mathcal {F}}_{h,1}}{\partial \psi _{2}}
- \frac {2\psi _{1}\psi _{2}^{2}}{\lambda _{1} \lambda _{2}}
\frac {\partial {\mathcal {F}}_{h,1}}{\partial \xi _{1}},
\nonumber \\
{\tilde {\mathcal {H}}}_{h,2}=
{\mathcal {H}}_{h,2} - \frac {\partial {\mathcal {P}}_{h,2}}{\partial \rho }
- \frac {\xi _{2}}{\lambda _{1}}
\frac {\partial {\mathcal {P}}_{h,2}}{\partial \psi _{2}},
\nonumber \\
{\mathcal {J}}_{h,1} =
-\frac {2\psi _{1}\psi _{2}^{2}}{\rho ^{2} \lambda _{1}}+
\frac {2\lambda _{2} \xi _{1}}{\rho ^{3}}
- \frac {\xi _{1}{\mathcal {G}}_{h,2}}{\rho ^{2}}.
\end{eqnarray}
We may now apply Theorem~\ref{thm:BFM}, which yields a four-parameter family of solutions in a local neighbourhood of $x=0$, of the form
\begin{eqnarray}
\rho = r_{h} + O(x), \qquad
{\tilde {\lambda }}_{1} = O(x),
\qquad
\lambda _{2} = S_{h} +O(x),
\nonumber \\
{\tilde {\psi }}_{1} = O(x), \qquad
\psi _{2} = \omega _{h} + O(x),
\nonumber \\
\xi _{1} = \frac {\alpha _{h}'r_{h}^{2}}{S_{h}} + O(x),
\qquad
{\tilde {\xi }}_{2} = O(x).
\end{eqnarray}
Restoring the original variables, we have proven the existence of solutions of the field equations in a neighbourhood of the event horizon, satisfying the boundary conditions (\ref{eq:horizon}) and analytic in $r_{h}$, $S_{h}$, $\alpha _{h}'$, $\omega _{h}$, $r$ and $\ell $.
\hfill $\square $

\paragraph{{\bf Comment}}

From (\ref{eq:Eeqns}) and the relation (\ref{eq:mu}), we can show that for the solutions described by Proposition~\ref{thm:horizon}, we have
\begin{equation}
\mu^\prime(r_h) = \frac{1}{r_h}\left(1+3\frac{r_h^2}{\ell^2}\right)
-\frac{r_h(\alpha_h')^2}{S_h^2}-\frac{1}{r_h^3}(\omega_h^2-1)^2.
\end{equation}
In order that $r>r_h$ corresponds to the \textit{exterior} of the black hole - i.e.\ to the untrapped region - we must have $\mu(r)>0$ for $r>r_h$.
Thus Proposition~\ref{thm:horizon} gives rise (locally) to black hole solutions when $\mu^\prime(r_h)>0$. This restricts the horizon data $(\ell,r_h,S_h,\omega_h,\alpha_h')$. However, for each choice of $r_h>0$ and $\ell$, there is an open set ${\cal{O}}$ of values of $(S_h,\omega_h,\alpha_h')$ in $\mathbb{R}^3$ that contains the horizon data $(S_h,\omega_h,\alpha_h')=(1,1,0)$ corresponding to the trivial Schwarzschild-adS solution and with the property that for each $(S_h,\omega_h,\alpha_h')\in {\cal{O}}$, Proposition~\ref{thm:horizon} describes (locally) a black hole solution. The existence of this open set ${\cal{O}}$ is crucial for the global existence result of the following section.

\begin{proposition}
\label{thm:infinity}
(Local existence of solutions in a neighbourhood of infinity) \\
There exists a five-parameter family of local solutions of the field equations
(\ref{eq:Eeqns}--\ref{eq:YMeqns}) near $r\rightarrow \infty $, satisfying the boundary conditions (\ref{eq:infinity}) and analytic in $M$, $\alpha _{\infty }$, $\omega _{\infty }$, $c_{1}$, $d_{1}$, $r^{-1}$ and $\ell $.
\end{proposition}

\paragraph{{\bf {Proof}}}
In this case we take $x=r^{-1}$ as our independent variable and define dependent variables as follows (again, these mostly follow \cite{BFM}):
\begin{eqnarray}
\lambda _{1}(x) = 2m(r) = r\left[ 1 - \mu (r) + \frac {r^{2}}{\ell ^{2}} \right] ,
\qquad
\lambda _{2}(x) = S(r),
\nonumber \\
\psi _{1}(x) = \alpha (r),
\qquad
\psi _{2}(x) = \omega (r),
\nonumber \\
\xi _{1}(x) = r^{2} \frac {d\alpha }{dr},
\qquad
\xi _{2}(x) = r^{2} \frac {d\omega }{dr}.
\end{eqnarray}
The field equations (\ref{eq:Eeqns}--\ref{eq:YMeqns}), when written in terms of our new variables, take the form
\begin{eqnarray}
x\frac {d\lambda _{1}}{dx} = x {\mathcal {G}}_{\infty ,1}
,
\qquad
x\frac {d\lambda _{2}}{dx} =x^{4} {\mathcal {G}}_{\infty ,2}
,
\qquad
x\frac {d\psi _{1}}{dx} =-x\xi _{1}
,
\nonumber \\
x\frac {d\psi _{2}}{dx} =-x\xi _{2}
,
\qquad
x\frac {d\xi _{1}}{dx} =x{\mathcal {H}}_{\infty ,1}
,
\qquad
x\frac {d\xi _{2}}{dx} =x{\mathcal {H}}_{\infty ,2}
,
\label{eq:infinityBFM}
\end{eqnarray}
where
\begin{eqnarray}
{\mathcal {F}}_{\infty } =\frac {1}{\ell ^{2}}+ x^{2} - x^{3}\lambda _{1},
\nonumber \\
{\mathcal {G}}_{\infty ,1} =
-1 - \frac {2\psi _{1}^{2} \psi _{2}^{2}}{{\mathcal {F}}_{\infty }\lambda _{2}^{2}}
+ \psi _{2}^{2} \left( 2 - \psi _{2}^{2} \right)
-\frac {\xi _{1}^{2}}{\lambda _{2}^{2}} - 2{\mathcal {F}}_{\infty }\xi _{2}^{2},
\nonumber \\
{\mathcal {G}}_{\infty ,2}=
-\frac {2\psi _{1}^{2}\psi _{2}^{2}}{{\mathcal {F}}_{\infty }^{2}\lambda _{2}}
- 2\lambda _{2}\xi _{2}^{2},
\nonumber \\
{\mathcal {H}}_{\infty ,1}=
-\frac {2\psi _{1}\psi _{2}^{2}}{{\mathcal {F}}_{\infty }}
+ \frac {x^{3}\xi _{1} {\mathcal {G}}_{\infty ,2}}{\lambda _{2}}
, \nonumber \\
{\mathcal {H}}_{\infty ,2} =
\frac {\psi _{1}^{2} \psi _{2}}{{\mathcal {F}}_{\infty }^{2} \lambda _{2}^{2}}
+ \frac {\psi _{2}\left( 1 - \psi _{2}^{2} \right) }{{\mathcal {F}}_{\infty }}
+\frac {x\xi _{2} \left( 3x\lambda _{1} - 2 \right) }{{\mathcal {F}}_{\infty }}
+ \frac {x^{3}\xi _{2} {\mathcal {G}}_{\infty ,1}}{{\mathcal {F}}_{\infty }}
- \frac {x^{3} \xi _{2}{\mathcal {G}}_{\infty ,2}}{\lambda _{2}}.
\nonumber \\
\label{eq:infquantities}
\end{eqnarray}
No further transformations are required as the equations (\ref{eq:infinityBFM}) are already in a form suitable for the application of Theorem~\ref{thm:BFM}.
The theorem then gives us a five-parameter family of local solutions in a neighbourhood of $x=0$, of the form
\begin{eqnarray}
\lambda _{1} = M +O(x), \qquad
\lambda _{2} = 1 + O(x^{4}),
\qquad
\psi _{1} = \alpha _{\infty } + O(x),
\nonumber \\
\psi _{2} = \omega _{\infty } + O(x),
\qquad
\xi _{1} = -d_{1} + O(x),
\qquad
\xi _{2} = -c_{1} + O(x).
\end{eqnarray}
Restoring the original variables, we therefore have proven the existence of solutions of the field equations in a neighbourhood of infinity, satisfying the boundary conditions (\ref{eq:infinity}) and analytic in $M$, $\alpha _{\infty }$, $\omega _{\infty }$, $c_{1}$, $d_{1}$,
$r^{-1}$ and $\ell $.
\hfill $\square $

\section{Global existence of black holes and solitons}
\label{sec:global}

In this section we turn to the issue of ``globalizing'' the existence and uniqueness results of the previous section. We consider first the black hole case and then, more briefly, the soliton case.

\subsection{Global existence proof for black holes}
\label{sec:BHglobal}

In order to construct asymptotically adS dyonic black hole space-times, we must prove the global existence of solutions to the EYM equations (\ref{eq:Eeqns}--\ref{eq:YMeqns}) with appropriate boundary conditions.
In this context, global existence refers to existence for all $r\geq r_h$, where $r_h$ is the horizon radius such that $\mu(r_h)=0$.
For each value of the adS radius of curvature $\ell$, the existence of a four-parameter family of such solutions is proven in this section using what is essentially a non-linear perturbation argument that relies on an application of the (Banach space) implicit function theorem - see, for example, \cite{deimling}.
See also \cite{hartman} for similar applications of this theorem to ODEs.

\begin{theorem}
\label{thm:IFT}
(Implicit function theorem) \\
Let $C,Y,Z$ be Banach spaces.
Let $U$ be a neighbourhood of ${\bf c_0}$ in  $C$, $V$ be a neighbourhood of
${\bf y_0}$ in $Y$ and let $F:U\times V\to Z$ be a (Fr\'echet) $C^1$ mapping.
Suppose that $F({\bf c_0},{\bf y_0})=0$ and that the linearization
$d_yF({\bf c_0},{\bf y_0})(0,\cdot):Y\to Z$ is an isomorphism.
Then there is a neighbourhood $\tilde{U}$ of ${\bf c_0}$ in $C$, a neighbourhood $\tilde{V}$ of ${\bf y_0}$ in $Y$ and a unique continuous mapping
$G:\tilde{U}\to \tilde{V}$ such that ${\bf y_0}=G({\bf c_0})$ and
$F({\bf c},G({\bf c}))=0$ for all $x\in \tilde{U}$.
\end{theorem}

Our aim then is to set up the field equations and boundary conditions in the form of a Banach space mapping.
In this approach, $C$ will correspond to the set (Banach space) of allowed initial values of the field variables, and $Y$ will correspond to the space of the field variables themselves  - a space of appropriately smooth functions.
The field equations are then written in a form that corresponds to a mapping $F$, with the condition that $F({\bf c},{\bf y})=0$ if and only if ${\bf y}$ satifies the field equations with initial values ${\bf c}$.
Then the trivial solution, Schwarzschild-adS, corresponds to a pair $({\bf c_0},{\bf y_0})\in C\times Y$ with $F({\bf c_0},{\bf y_0})=0$.
The existence of the mapping $G$ of the theorem then gives rise to non-trivial solutions ${\bf y}=G({\bf c})$ of the Banach space mapping, which will correspond to non-trivial solutions $F({\bf c},G({\bf c}))=0$ of the field equations and initial conditions.

The key to applying the implicit function theorem is to write the field equations in a form in which the linearization about the trivial solution has certain nice properties throughout the interval $[r_{h},+\infty)$.
However, this is essentially impossible, as the linearized equations are singular at $r=r_{h}$.
To avoid this difficulty, we use the local existence results at the horizon
(Proposition~\ref{thm:horizon}) to construct solutions on an interval of the form $[r_{h},r_{h}+\delta]$ where $\delta>0$.
Then, with a suitable choice of dependent and independent variables, the linearized field equations do have the required properties throughout $[r_h+\delta,+\infty)$, and the implicit function theorem can be applied to produce solutions on this interval. This is carried out in Lemma~\ref{prop:globalq}. These are a continuation of the local solutions, and so overall we have solutions on $[r_h,+\infty)$ - that is, global solutions. The proof of this continuation is given in Proposition~\ref{thm:BHglobal} and Lemma~\ref{lem:manifold}.

We begin by considering again the formulation of the field equations used for the proof of local existence at infinity (see Proposition~\ref{thm:infinity}). We recall that $x=r^{-1}$ and we use the notation
\begin{equation}
{\bf y}=\left(
2m, S, \alpha, \omega, r^2\frac{d\alpha}{dr}, r^2\frac{d\omega}{dr}
\right)^T.
\label{eq:ydef}
\end{equation}
Then the equations (\ref{eq:Eeqns}--\ref{eq:YMeqns}) take the form
\begin{equation}
\frac{d{\bf y}}{dx}
={\bf A}(x,{\bf y})
=\left(
{\mathcal {G}}_{\infty ,1},x^3{\mathcal {G}}_{\infty ,2},-y_5,-y_6,{\mathcal {H}}_{\infty ,1},{\mathcal {H}}_{\infty ,2}
\right)^T.
\end{equation}
The coefficients on the right hand side are given in the proof of
Proposition~\ref{thm:infinity} (\ref{eq:infquantities}).
If
\begin{equation}
f={\mathcal {F}}_{\infty}=x^2-y_1 x^3+\frac{1}{\ell^2}=\frac{\mu}{r^2}
\end{equation}
 and $y_2=S$ are non-zero, this system of equations is equivalent to
\begin{eqnarray}
Q_1:=fy_2^2\frac{dy_1}{dx}+H_1(x,{\bf y})=0,
\nonumber \\
Q_2:=f^2y_2\frac{dy_2}{dx}+H_2(x,{\bf y})=0,
\nonumber \\
Q_3:=\frac{dy_3}{dx}+H_3(x,{\bf y})=0,
\nonumber \\
Q_4:=\frac{dy_4}{dx}+H_4(x,{\bf y})=0,
\nonumber \\
Q_5:=f^2y_2^2\frac{dy_5}{dx}+H_5(x,{\bf y})=0,
\nonumber \\
Q_6:=f^2y_2^2\frac{dy_6}{dx}+H_6(x,{\bf y})=0,
\label{eq:qeqns}
\end{eqnarray}
where
\begin{eqnarray}
H_1(x,{\bf y})
&=&
 fy_2^2+2y_3^2y_4^2-fy_2^2y_4^2(2-y_4^2)+fy_5^2+2f^2y_2^2y_6^2,
 \nonumber\\
H_2(x,{\bf y})
&=&
 2x^3(y_3^2y_4^2+f^2y_2^2y_6^2),
 \nonumber\\
H_3(x,{\bf y})
&=&
y_5,
\nonumber\\
H_4(x,{\bf y})
&=&
 y_6,
 \nonumber\\
H_5(x,{\bf y})
&=&
 2f y_2^2 y_3 y_4^2+y_5H_2,
 \nonumber\\
H_6(x,{\bf y})
&=&
 -y_3^2y_4-fy_2^2y_4(1-y_4^2)-xfy_2^2(3xy_1-2)y_6+y_6(x^3H_1-H_2).
 \nonumber \\
\end{eqnarray}
We note that the functions $H_i, 1\leq i\leq 6$ are analytic for all $x\in\mathbb{R}$ and ${\bf y}\in\mathbb{R}^6$.

Let $x_*>\frac{1}{r_h}$. We now introduce the Banach spaces $C=\mathbb{R}^6$, $Y=C^1([0,x_*],\mathbb{R}^6)$ and $Z=\mathbb{R}^6\times C^0([0,x_*],\mathbb{R}^6)$ and define the mapping
\begin{equation}
F:C\times Y\to Z \quad:\quad
({\bf c},{\bf y})\in C\times Y \mapsto F({\bf c},{\bf y})
= \left( \begin{array}{c}
{\bf y}(x_*)-{\bf c} \\
{\bf Q}(x,{\bf y})
\end{array}\right),
\label{eq:mapping}
\end{equation}
where
${\bf {Q}}(x,{\bf {y}}) = \left( Q_{1}, \ldots , Q_{6} \right)^{T}$.
Note that $F({\bf c},{\bf y})=0$ if and only if ${\bf y}$ is a solution of the ODEs (\ref{eq:qeqns}) satisfying the boundary conditions ${\bf y}(x_*)={\bf c}$. We define
\begin{equation}
{\bf c_0}={\bf y_0}=\left( 2m_0,1,0,1,0,0 \right) ^T.
\end{equation}
These values correspond to the trivial solution
\begin{equation}
\mu = 1-\frac{2m_0}{r}+\frac{r^2}{\ell^2},\quad S=1,\quad\alpha=0,\quad\omega=1
\end{equation}
of the field equations (and so $F({\bf c_0},{\bf y_0})=0$).
The linearization $d_yF$ of $F$ at $({\bf c_0},{\bf y_0})$ is
\begin{equation}
d_yF({\bf c_0},{\bf y_0})(0,{\bf v})
=\left(
\begin{array}{c}
{\bf v}(x_*) \\
\delta {\bf Q}(x,{\bf y_0},{\bf v}))
\end{array}
\right),
\label{eq:lindef}
\end{equation}
where
\begin{equation}
\delta Q_i(x,{\bf y_0},{\bf v})=
\frac{\partial Q_i}{\partial p_i}
(x,{\bf y_0})
\frac{dv_i}{dx}
+\sum_{j=1}^6\frac{\partial Q_i}{\partial y_j}(x,{\bf y_0})v_j,
\end{equation}
with $p_i=\frac{d y_i}{dx}$. A straightforward calculation yields
\begin{eqnarray}
\delta Q_1 = f_0\frac{dv_1}{dx},
\nonumber\\
\delta Q_2 = f_0^2\frac{dv_2}{dx},
\nonumber\\
\delta Q_3 = \frac{dv_3}{dx}+v_5,
\nonumber\\
\delta Q_4 = \frac{dv_4}{dx}+v_6,
\nonumber\\
\delta Q_5 = f_0^2\frac{dv_5}{dx}+2f_0v_3,
\nonumber\\
\delta Q_6 = f_0^2\frac{dv_6}{dx}+2f_0v_4+2x(1-3m_0x)f_0v_6,
\label{eq:delqform}
\end{eqnarray}
where
\begin{equation}
f_0 = x^2-2m_0x^3+\frac{1}{\ell^2}.
\end{equation}
Note that by the definition of $x_*$ and $r_h$, we have $f_0(x)>0$ for all $x\in[0,x_*]$.

The next step is to determine if the mapping
\begin{equation}
d_yF({\bf c_0},{\bf y_0})(0,\cdot)\quad:\quad{\bf v}\in Y \mapsto
d_yF({\bf c_0},{\bf y_0})(0,{\bf v})\in Z
\end{equation}
is an isomorphism.
This is the case if for every ${\bf d}\in \mathbb{R}^6$ and every ${\bf z}\in C^0([0,x_*],\mathbb{R}^6)$ there exists a unique
${\bf v}\in C^1([0,x_*],\mathbb{R}^6)$ such that
\begin{equation}
{\bf v}(x_*)={\bf d},\qquad \delta {\bf Q}(x,{\bf y_0},{\bf v})={\bf z}.
\end{equation}
This follows immediately by the standard theory of linear ordinary differential equations \cite{codlev}.
The linear equations $\delta {\bf Q} = {\bf z}$ are regular throughout the interval $[0,x_*]$ and the terminal conditions ${\bf v}(x_*)={\bf d}$ pick out a unique solution.
The fact that $f_0>0$ throughout $[0,x_*]$ is crucial for this argument. Thus applying the implicit function theorem yields the following.

\begin{lemma}
\label{prop:globalq}
There is a neighbourhood $U$ of ${\bf c_0}\in \mathbb{R}^6$
and a neighbourhood $V$ of ${\bf y_0}\in C^1([0,x_*],\mathbb{R}^6)$ such that for each ${\bf c}\in U$, there is a solution ${\bf y}\in V$
of the equations (\ref{eq:qeqns}) on $[0,x_*]$ with ${\bf y}(x_*)={\bf c}$.
The neighbourhoods $U$ and $V$ may be chosen so that $\mu>0$ and $S>0$ throughout the interval $[0,x_*]$ and there is a unique ${\bf y}\in V$ for each ${\bf c}\in U$.
\end{lemma}

\paragraph{{\bf {Proof}}}

We take the Banach spaces $C,Y,Z$ and the mapping $F$ as in the preceding paragraphs. Then Theorem~\ref{thm:IFT} yields neighbourhoods $\tilde{U}, \tilde{V}$ of
${\bf c_0}, {\bf y_0}$ respectively and a unique continuous mapping $G:\tilde{U}\to\tilde{V}$ with
$G({\bf c_0})={\bf y_0}$ and $F({\bf c},G({\bf c}))=0$ for all ${\bf c}\in\tilde{U}$. By the definition of $F$, each ${\bf y}=G({\bf c})$ is a $C^1$ solution of (\ref{eq:qeqns}) on $[0,x_*]$ that satisfies the condition ${\bf y}(x_*)={\bf c}$. Next, consider an $\epsilon-$ball $B_\epsilon({\bf y_0})\subset \tilde{V}$, and let ${\bf y}\in B_\epsilon({\bf y_0})$.
Then (using the standard maximum norm for $C^1[0,x_*]$)
\begin{equation}
\left| y_1-y_1({\bf c_0}) \right| <\epsilon \quad \hbox{for all } x\in[0,x_*],
\end{equation}
that is,
\begin{equation}
\left| 2m-2m_0 \right| <\epsilon \quad \hbox{for all } r\geq r_*>r_h.
\end{equation}
It follows that
\begin{equation}
\mu(r) = 1-\frac{2m}{r}+\frac{r^2}{\ell^2}>1-\frac{2m_0+\epsilon}{r}+\frac{r^2}{\ell^2}\quad \hbox{for all } r\geq r_*>r_h.
\end{equation}
Recalling that
\begin{equation}
\mu_0(r_h) = 1-\frac{2m_0}{r_h}+\frac{r_h^2}{\ell^2}=0,
\end{equation}
it is straightforward to show that a sufficiently small $\epsilon>0$ can always be chosen so that
\begin{equation}
1-\frac{2m_0+\epsilon}{r}+\frac{r^2}{\ell^2}>0\quad \hbox{for all } r\geq r_*>r_h,
\end{equation}
from which it follows that $\mu>0$ throughout $[0,x_*]$.
A similar, but more straightforward argument shows the same conclusion for $S$.
We now take $V=B_\epsilon({\bf y_0})$ for the sufficiently small positive $\epsilon$ found above, and take $U=G^{-1}(V)$.
By continuity, this is open (and must contain ${\bf c_0}$).
Positivity of $\mu$ and $S$ show that $f=x^2-x^3u_1+\ell^{-2}$ and $y_2=S$ are bounded away from zero on the interval $[0,x_*]$.
Then uniqueness of ${\bf y}\in{V}$ as a solution of (\ref{eq:qeqns}) for a given
${\bf c}\in{U}$ follows by the standard argument of estimating the growth of
$\delta {\bf y}$, the difference of two potentially distinct solutions.
(It is crucial to have bounds on the coefficients of the derivatives in (\ref{eq:qeqns}) for this standard argument to apply; these coefficients are powers of $f$ and $y_2$.)
\hfill$\square$

\paragraph{{\bf {Comment}}}
Lemma~\ref{prop:globalq} establishes the existence of solutions of (\ref{eq:qeqns}) on the interval $[0,x_*]$. These solutions have $f=\mu/r^2>0$ and $y_2=S>0$, and so correspond to solutions of (\ref{eq:Eeqns}--\ref{eq:YMeqns}), defined on an interval of the form $[r_*,+\infty)$ where $r_*>r_h$. In the following results, we prove that an open subset of these solutions provide a continuation of the local solutions at the horizon, and thus yield global solutions defined for all $r\geq r_h$. This proof relies on the geometric structure of the set of local solutions near the horizon.

\begin{proposition}
\label{thm:BHglobal}
(Global existence of black hole solutions) \\
Let $r_h>0, \Lambda<0$ and define $\frac{1}{\ell^2}=-\frac{\Lambda}{3}$.
Then there is an open neighbourhood $\Gamma^{\rm bh}$ of ${\bf g_0}=(1,1,0)\in \mathbb{R}^3$ such that for each
${\bf g}=(S_h,\omega_h,\alpha_h')\in\Gamma^{\rm bh}$, there exists a global solution of the EYM field equations
(\ref{eq:Eeqns}--\ref{eq:YMeqns}) defined on $[r_h,+\infty)$ with the asymptotic behaviour (\ref{eq:horizon})
as $r\rightarrow r_{h}^{+}$ and the asymptotic behaviour (\ref{eq:infinity}) as $r\rightarrow  \infty.$
These solutions are differentiable on $[r_h,\infty)$ ($C^1$ for $\mu,S$ and $C^2$ for $\alpha,\omega$) and satisfy $\mu(r)>0$ for all $r>r_h$. Since $\mu(r_h)=0$, the region $r>r_h$ corresponds to the exterior of a black hole.
\end{proposition}

\paragraph{{\bf {Proof}}}

By Proposition~\ref{thm:horizon} and the comment which follows its proof, there is a neighbourhood $\Gamma$ of ${\bf g_0}\in\mathbb{R}^3$ such that for each ${\bf g}\in\Gamma$, we obtain a solution of the EYM field equations
(\ref{eq:Eeqns}--\ref{eq:YMeqns}) on an interval $[r_h,r_h+\delta)$ corresponding locally to a black hole solution.
These solutions are analytic in $r$ and in ${\bf g}$ on this interval.
Now choose $x_*=r_*^{-1}$ with $r_h<r_*<r_h+\delta$.
Then it follows by the same proposition that the solution vector ${\bf y}$ defined by (\ref{eq:ydef}) is analytic in ${\bf g}$ at $x=x_*$.
Let ${\bf y}(x;{\bf g})$ denote the solution ${\bf y}(x)$ with horizon data ${\bf g}$ and define a mapping
\begin{equation}
\phi:\Gamma\to \mathbb{R}^6  \quad : \quad {\bf g}\in\Gamma\mapsto\phi({\bf g})
={\bf y}(x_*;{\bf g}).
\label{eq:phidef}
\end{equation}
Note that $\phi$ is analytic on $\Gamma$ as a consequence of
Proposition~\ref{thm:horizon}. The image
${\mathcal {M}}$ of $\phi$,
\begin{equation}
{\mathcal {M}} =\phi(\Gamma) =
\{ \phi({\bf g})\in\mathbb{R}^6:{\bf g}\in\Gamma\}
\end{equation}
is a 3-parameter subset of $\mathbb{R}^6$ that contains the point ${\bf c_0}=(2m_0,1,0,1,0,0)$.
(Recall that this vector ${\bf c_0}\in\mathbb{R}^6$ corresponds to the trivial solution of the field equations with horizon data ${\bf g_0}$.)
We can show that without loss of generality, ${\mathcal {M}}$ is a three-dimensional submanifold of $\mathbb{R}^6$ in an neighbourhood of ${\bf c_0}$:
this is proven in Lemma~\ref{lem:manifold} below.
Now define $\hat{U}={\mathcal {M}}\cap U$, where $U$ is the neighbourhood of
${\bf c_0}$ described by Lemma~\ref{prop:globalq}. By this lemma,
for each ${\bf c}\in\hat{U}$, there is a unique solution of the EYM equations on $[r_*,\infty)$.
Since ${\mathcal {M}}$ is a manifold, it follows that $\hat{U}$ is an open subset of ${\mathcal {M}}$.
Thus each ${\bf g}\in\Gamma^{\rm bh}$ gives rise to a solution of the EYM equations on $[r_h,+\infty)$, where
\begin{equation}
\Gamma^{\rm bh} = \phi^{-1}(\hat{U}) =
\{{\bf g}\in\Gamma:{\bf y}(x_*;{\bf g})\in \hat{U}\}
\end{equation}
is an open subset of $\Gamma$, being the inverse image of an open set with respect to an analytic (and hence continuous) mapping.
The asymptotics carry over from the corresponding local results at the horizon and at infinity respectively.
\hfill$\square$

\begin{lemma}
\label{lem:manifold}
Let $\Gamma$ and $\phi$ be as in the proof of
Proposition~\ref{thm:BHglobal}. Then there is an open subset $\Gamma_1$ of $\Gamma$ containing ${\bf g_0}$ such that ${\mathcal {M}}_1=\phi(\Gamma_1)$ is a three-dimensional submanifold of $\mathbb{R}^6$.
\end{lemma}

\paragraph{{\bf {Proof}}}
Let us write ${\bf g}=(\xi,\eta,\zeta)=(S_h,\omega_h,\alpha_h')\in\Gamma$ for the horizon data, and write
\begin{equation}
{\mathcal {M}}=\{\phi(\xi,\eta,\zeta)\in\mathbb{R}^6: (\xi,\eta,\zeta)\in\Gamma\}.
\end{equation}
We show that in a neighbourhood of $\phi({\bf g_0})$, we can describe ${\mathcal {M}}$ as the graph of a $C^1$ function. To see this, we recall that the local solution described by Proposition~\ref{thm:horizon} is analytic in both $r$ and the horizon data ${\bf g}$ for values of $r$ sufficiently close to $r_h$.
The quantity $x_*$ corresponds to a value in this range, and so $\phi$ is analytic in ${\bf g}$.
Recalling the definition (\ref{eq:ydef}) and the asymptotics derived as part of Proposition~\ref{thm:horizon}, we have
\begin{eqnarray}
\phi_2 = \xi + O(r-r_h),
\nonumber \\
\phi_4 = \eta + O(r-r_h),
\nonumber \\
\phi_5 = r_h^2\zeta + O(r-r_h).
\end{eqnarray}
In fact, by analyticity, the remainder terms can be replaced by convergent power series in $r-r_h$, and so may be differentiated with respect to $\xi,\eta$ and $\zeta$. That is, we can apply the rule
\[ \frac{\partial}{\partial \xi}\left[O(r-r_h)\right]=O(r-r_h),\]
and the corresponding results for $\eta$ and $\zeta$. The coefficients of these power series are analytic functions of $(\xi,\eta,\zeta)$. Thus the Jacobian matrix of the mapping $\bar{\phi}:(\xi,\eta,\zeta)\to(\phi_2,\phi_4,\phi_5)$ has the form
\begin{equation}
\frac{\partial(\phi_2,\phi_4,\phi_5)}{\partial(\xi,\eta,\zeta)}=
\left(
\begin{array}{ccc}
         1+O(r-r_h) & O(r-r_h) & O(r-r_h) \\
         O(r-r_h) & 1+O(r-r_h) & O(r-r_h) \\
         O(r-r_h) & O(r-r_h) & r_h^2+O(r-r_h)
       \end{array}\right),
\end{equation}
with determinant
\begin{equation}
J(\bar{\phi}) = \left| \frac{\partial(\phi_2,\phi_4,\phi_5)}{\partial(\xi,\eta,\zeta)} \right|= r_h^2+O(r-r_h).
\end{equation}
Thus for $r_*$ chosen sufficiently close to $r_h$, this determinant is non-zero and the inverse function theorem \cite{deimling} shows that there exist $C^1$ functions $\psi_1,\psi_2,\psi_3$ and an open neighbourhood $\Gamma_1$ of $(1,1,0)$ such that
\begin{equation}
\left( \begin{array}{c}
            y_2 \\
            y_4 \\
            y_5
          \end{array}
          \right)=\left( \begin{array}{c}
            \phi_2(\xi,\eta,\zeta) \\
            \phi_4(\xi,\eta,\zeta) \\
            \phi_5(\xi,\eta,\zeta)
          \end{array}\right)
          \Leftrightarrow
          \left( \begin{array}{c}
            \xi \\
            \eta \\
            \zeta
          \end{array}\right)=
          \left( \begin{array}{c}
            \psi_1(y_2,y_4,y_5) \\
            \psi_2(y_2,y_4,y_5) \\
            \psi_3(y_2,y_4,y_5)
          \end{array}\right) ,
\end{equation}
for $(\xi,\eta,\zeta)\in\Gamma_1$ and $(y_2,y_4,y_5)\in\bar{\phi}(\Gamma_1)$.
Now define ${\mathcal {M}}_1=\phi(\Gamma_1)$.
This can be written as
\begin{equation}
{\mathcal {M}}_1 = \left\{
\left( \begin{array}{c}
            y_1 \\
            y_2 \\
            y_3 \\
            y_4 \\
            y_5 \\
            y_6
          \end{array}\right)=\left( \begin{array}{c}
            \tilde{\phi}_1(y_2,y_4,y_5) \\
            y_2 \\
            \tilde{\phi}_3(y_2,y_4,y_5) \\
            y_4 \\
            y_5 \\
            \tilde{\phi}_6(y_2,y_4,y_5)
          \end{array}\right)
          : (y_2,y_4,y_5)\in \bar{\phi}(\Gamma_1)
          \right\},
\end{equation}
where $\tilde{\phi}_i=\phi_i\circ\psi$ - that is, $\tilde{\phi}_i(y_2,y_4,y_5)
=\phi_i(\psi_1(y_2,y_4,y_5),\psi_1(y_2,y_4,y_5),\psi_1(y_2,y_4,y_5))
$,
$i=1,3,6$.
This shows that ${\mathcal {M}}_1$ is the graph of a smooth ($C^1$) function, and is therefore (counting dimensions) a three-dimensional submanifold of $\mathbb{R}^6$ - again, by an application of the inverse function theorem. See, for example, chapter 3 of \cite{darling} for further details of this last point.
\hfill$\square$

\paragraph{{\bf {Comment}}}

Proposition~\ref{thm:BHglobal} has proven the existence of non-trivial black hole solutions of the field equations (\ref{eq:Eeqns}--\ref{eq:YMeqns}) in a neighbourhood
of the trivial Schwarzschild-adS solution.
The proof is valid for any value of the adS radius of curvature $\ell $, but does
not give us any information about the size of the neighbourhood in which we have non-trivial solutions.
Numerical investigations \cite{Bjoraker,Shepherd} indicate that the size of this
neighbourhood will shrink as $\ell $ increases (or, as the magnitude of the
cosmological constant $\left| \Lambda \right| $ decreases).
Since the results of section~\ref{sec:local} tell us that the solutions are analytic
in the horizon data, non-trivial solutions which are sufficiently close to the trivial solution will be such that the magnetic gauge field function $\omega (r)$ has
no zeros since the trivial solution $\omega (r) \equiv 1$ has no zeros.
We anticipate that at least some of these solutions will be stable.


\subsection{Global existence proof for solitons}
\label{sec:sol-global}

The method applied above to establish the global existence of asymptotically adS black holes carries over with minimal changes to the soliton case.
We highlight those changes here and then quote the relevant result.

The argument proceeds unchanged up to and including the definition (\ref{eq:mapping}) of the Banach space mapping $F$.
The linearization is now about the trivial data point and trivial solution
\begin{equation}
{\bf c_0}={\bf y_0}=(0,1,0,1,0,0)^T
\end{equation}
corresponding to pure anti-de Sitter space-time wherein
\begin{equation}
\mu(r)=1+\frac{r^2}{\ell^2},\quad S(r)=1,\quad \alpha(r)=0,\quad \omega(r)=1.
\end{equation}
The linearization (\ref{eq:lindef}) maintains the same form (\ref{eq:delqform}) but with $m_0=0$, and hence $f_0=x^2+\ell^{-2}$.
Lemma~\ref{prop:globalq} thus carries through. In the present case, $x_*$ is the reciprocal of a value $r_*$ with $0<r_*<\delta$, where the local existence result Proposition~\ref{thm:origin} applies on $[0,\delta)$. These local solutions satisfy the boundary conditions (\ref{eq:originfinal}), which are equivalent to
\begin{eqnarray}
y_1=2m =2m_3r^3+O(r^4),
\quad y_2=S=S_0+S_2r^2+O(r^3),
\nonumber\\
y_3=\alpha=\alpha_1r+\alpha_3r^3+O(r^4),
\quad y_4=\omega = 1+\omega_2r^2+O(r^3),
\nonumber\\
y_5=r^2\alpha'=\alpha_1r^2+O(r^3),
\quad y_6=r^2\omega' = 2\omega_2r^3+O(r^4).
\label{eq:ysolbcs}
\end{eqnarray}
The free parameters here are $(\xi,\eta,\zeta)=(S_0,\alpha_1,\omega_2)$, with the other parameters being given in terms of these (see (\ref{eq:originfinal})).
These represent the free initial data at the origin for solitons, and local existence of an analytic solution applies for data in a neighbourhood $\Gamma$ of the trivial data $(\xi,\eta,\zeta)=(1,0,0)$.
As in the black hole case, the local existence theorem gives rise to an analytic mapping defined as in (\ref{eq:phidef}).
Corresponding to $\bar{\phi}$, we define the mapping $\hat{\phi}:(\xi,\eta,\zeta)\to(\phi_2,\phi_3,\phi_4)$, which has Jacobian matrix of the form
\begin{equation}
\frac{\partial(\phi_2,\phi_3,\phi_4)}{\partial(\xi,\eta,\zeta)}=
\left(
\begin{array}{ccc}
         1+O(r^2) & O(r^2) & O(r^2) \\
         O(r^3) & r+O(r^3) & O(r^3) \\
         O(r^3) & O(r^3) & r^2+O(r^3)
       \end{array}
       \right),
\end{equation}
with determinant
\begin{equation}
J(\hat{\phi}) =
\left| \frac{\partial(\phi_2,\phi_3,\phi_4)}{\partial(\xi,\eta,\zeta)} \right|= r^3+O(r^6).
\end{equation}
Thus for sufficiently small $r$, we see that $\hat{\phi}$ is a diffeomorphism in a neighbourhood of the trivial data.
This is the last technical result required to ensure that the analogues of
Proposition~\ref{thm:BHglobal} and Lemma~\ref{lem:manifold} carry through to the soliton case, and so we can quote the following result.

\begin{proposition}
\label{thm:SOLglobal}
(Global existence of soliton solutions) \\
Let $\Lambda<0$ and define $\frac{1}{\ell^2}=-\frac{\Lambda}{3}$.
Then there is an open neighbourhood $\Gamma^{\rm sol}$ of
${\bf g_0}=(1,0,0)\in \mathbb{R}^3$ such that for each
${\bf g}=(S_0,\alpha_1,\omega_2)\in\Gamma^{\rm sol}$, there exists a global solution of the EYM field equations (\ref{eq:Eeqns}--\ref{eq:YMeqns}) defined on $[0,+\infty)$ with the asymptotic behaviour (\ref{eq:originfinal}) as $r\to 0$ and the asymptotic behaviour (\ref{eq:infinity}) as $r\to+\infty$.
These solutions are differentiable ($C^1$ for $\mu$ and $S$, $C^2$ for $\alpha,\omega$) on $[0,\infty)$ and satisfy $\mu(r)>0$ for all $r\geq 0$.
\end{proposition}

\paragraph{{\bf {Comment}}}

Proposition~\ref{thm:SOLglobal} has proven the existence of non-trivial soliton solutions of the field equations (\ref{eq:Eeqns}--\ref{eq:YMeqns}) in a neighbourhood
of the trivial pure adS solution.
As with the corresponding result for black holes, the proof is valid for any value of the adS radius of curvature $\ell $, but does
not give us any information about the size of the neighbourhood in which we have non-trivial solutions.
For solitons as well as black holes, numerical work \cite{Bjoraker,Shepherd1} shows that the size of this
neighbourhood will shrink as $\ell $ increases.
Since the results of section~\ref{sec:local} tell us that the solutions are analytic
in the initial data at the origin, non-trivial solutions which are sufficiently close to the trivial solution will be such that the magnetic gauge field function $\omega (r)$ has
no zeros, solutions which are of particular interest because we expect at least some of them to be stable.


\section{Conclusions}
\label{sec:conc}

In this paper we have studied dyonic soliton and black hole solutions of the
${\mathfrak {su}}(2)$ EYM equations in adS.  These solutions have been found numerically some years ago \cite{Bjoraker} but, while the purely magnetic system has been studied in depth analytically, the dyonic system has received comparatively little attention in the literature.
Here we have proven analytically the existence of solutions of the field equations representing dyonic solitons and black holes.
The approach is two-fold: firstly, we prove the local existence of solutions near the origin, black hole event horizon and infinity, where the governing differential equations are singular.
Global existence of solutions is then proven using a non-linear perturbation argument which relies on an application of the Banach space implicit function theorem.  This approach to proving global existence of solutions of the EYM equations in adS is novel - previous work (for example \cite{Winstanley1}) has used a ``shooting''-type approach based on showing that the local solutions in a neighbourhood of the origin or event horizon can be extended to large values of $r$, and then matching onto the local solutions in a neighbourhood of infinity.


The result proved in this paper is valid for any non-zero adS radius of curvature
$\ell $ (or, equivalently, any negative cosmological constant $\Lambda <0$).
Since it is known \cite{Ershov} that there are no non-trivial dyonic solutions of the EYM field equations when the space-time is asymptotically flat rather than
asymptotically adS, a natural question is how our proof breaks down in the
$\Lambda =0$, $\ell \rightarrow \infty $ case.
The key difference between the asymptotically flat and asymptotically adS cases is the boundary conditions as $r\rightarrow \infty $.
In particular, Proposition \ref{thm:infinity} (giving local existence of solutions near infinity) does not hold as currently stated because ${\mathcal {F}}_{\infty }=0$ when $x=0$ and $\ell \rightarrow \infty $, so the ordinary differential equations
(\ref{eq:infinityBFM}) are not regular.
This result breaks down because, in asymptotically flat space, the boundary conditions at infinity imply that either $\alpha $ or $\omega $ must vanish as $r\rightarrow \infty $.


The global existence result in this paper proves the existence of non-trivial solutions in a neighbourhood of the trivial solution $\alpha (r) \equiv 0$, $\omega (r) \equiv 1$, $M(r) \equiv {\mathrm {constant}}$, $S(r)\equiv 1$, corresponding to either pure adS or the Schwarzschild-adS black hole.
In particular, at least some of these solutions, sufficiently close to the trivial solution, are such that the magnetic gauge function $\omega (r)$ has no zeros.
These solutions are particularly interesting because one may conjecture, in analogy with the purely magnetic case \cite{Winstanley1,Bjoraker} that at least some such solutions will be linearly stable.
We will return to this question in a forthcoming publication.


It might also be of interest to investigate whether the analytic approach taken in this paper can be applied to soliton and black hole solutions of Einstein gravity coupled to other matter fields.
For example, adding scalar fields to the model (possibly with coupling to the gauge field) would be relevant to supergravity models with non-abelian gauge fields \cite{SUGRA}.
We leave such questions for future research.

\ack

We acknowledge support from the Office of the Vice President for Research in Dublin City University for an International Visitor Programme grant which enabled the completion of this work.
B.N. thanks Fabio Giannoni and Roberto Giamb\'o for useful discussions, several years ago, on the implicit function theorem.
The work of E.W. is supported by the Lancaster-Manchester-Sheffield
Consortium for Fundamental Physics under STFC grant ST/J000418/1 and by EU COST Action MP0905 ``Black Holes in a Violent Universe''.
E.W. thanks Dublin City University for hospitality while this work was in progress. E.W. thanks Ben Shepherd for useful discussions.

\end{document}